\magnification=\magstep1
\input iopppt.modifie
\input xref
\input epsf
\eqnobysec

%
\def\received#1{\insertspace 
     \parindent=\secindent\ifppt\textfonts\else\smallfonts\fi 
     \hang{Received #1}\rm } 
\headline={\ifodd\pageno{\ifnum\pageno=\firstpage\titlehead
   \else\rrhead\fi}\else\lrhead\fi} 
\def\figure#1{\global\advance\figno by 1\gdef\labeltype{\figlabel}%
   {\parindent=\secindent\smallfonts\hang 
    {\bf Figure \ifappendix\applett\fi\the\figno.} \rm #1\par}} 
\footline={\ifnum\pageno=\firstpage
\hfil\textfonts\folio\fi}   
\def\titlehead{\smallfonts J. Phys. A: Math. Gen. {\bf 33} (2000)
2587--2595
\hfil}

\firstpage=2587
\pageno=2587


\jnlstyle

\jl{1}

\title{Lattice animals on a staircase and Fibonacci
numbers}[Lattice animals on a staircase and Fibonacci numbers]

\author{L Turban}[L Turban]

\address{Laboratoire de Physique des 
Mat\'eriaux, Universit\'e Henri Poincar\'e,
BP~239,  F--54506~Vand\oe uvre l\`es Nancy Cedex, France}

\received{16 November 1999}

\abs
We study the statistics of column-convex lattice animals resulting from
the stacking of squares on a single or double staircase. We obtain exact
expressions for the number of animals with a given length and area, their
mean length and their mean height. These objects are closely related to
Fibonacci numbers. On a single staircase, the total number of animals with
area $k$ is given by the Fibonacci number~$F_k$.   
\endabs



\submitted

\date

\section{Introduction}
A {\it lattice animal} is a cluster of occupied sites on a lattice. In two
dimensions, the {\it area} of an animal is the number of sites belonging the
cluster and its {\it perimeter} is defined as the set of empty
first neighbours of ocupied sites. Alternatively, instead of the site
clusters, one may consider the corresponding clusters of occupied cells on
the dual lattice, also called {\it polyominoes}.    

The enumeration of lattice animals according to their
area and/or perimeter is a subject of active research in the statistical
physics and combinatorics communities (see~\cite{bousquet96a} for a recent
review). On the physical side, the main interest lies in the close
connection between lattice animals and the percolation
problem~\cite{stauffer92,bousquet96b}. 

In the most general case, the counting problem is quite difficult and only
some bounds on the asymptotic behaviour are known~\cite{klarner73}. This led
to the introduction of restricted classes of animals (Ferrers graphs,
convex and/or directed animals) for which some exact results could be
obtained, mainly in two dimensions (see for example~[5--12]).

In the present work, we consider animals resulting from the
stacking of squares on a single or double staircase, i.e. column-convex
(or vertically convex) animals, for which the intersection of a vertical line
with the perimeter has at most two connected components. These animals
are closely related to Fibonacci numbers. In the case of
a single staircase, the correspondance is particularly simple: the
area-generating function is {\it equal} to the Fibonacci number generating
function.  

The paper is organized as follows. In section 2, we study the stacking of
squares on a single staircase. We calculate the number of animals with a
given length and area, the total number of animals with a given area, their
mean length and mean height and study the asymptotic behaviours. The same
is done in section 3 for animals on a double staircase. The results are
discussed in section 4. Some technical details are given in the appendix.

\section{Single staircase}
In this section we consider column-convex lattice animals resulting from the
stacking of $k$ squares on a single staircase as shown in figure~\ref{fig1}.
Two neighbouring columns are connected when they share at least one edge. 

\subsection{Number of animals}
In order to count the number $F_{k,l}$ of different animals with area $k$ 
living on $l$ stairs, we introduce the generating function
$$
F(z,t)=\sum_{k,l=1}^\infty F_{k,l}\, z^kt^l\,.
\label{e2.1}
$$
It satisfies the relation
$$
F(z,t)=zt+t\,{z^2\over1-z}\,[1+F(z,t)]
\label{e2.2}
$$
where the first term corresponds to a single square, the factor $z^2/(1-z)$
in the second term is the generating function of a column with at least two
squares, needed for the animal to eventually continue its growth on
the next stair. Hence we have 
$$
\eqalign{
F(z,t)&={zt\over1-z-tz^2}=zt\sum_{n=0}^\infty z^n(1+zt)^n\cr
&=\sum_{n=0}^\infty\sum_p{n\choose p}z^{n+p+1}t^{p+1}
=\sum_l\sum_{k=l}^\infty{k-l\choose l-1}z^kt^l\,.}
\label{e2.3} 
$$
By convention, in sums containing binomial coefficients, the range of
summation is not explicitly indicated. It is automatically
determined by the nonvanishing values of the binomial coefficients and, for
example, $0\leq p\leq n$ in equation~\ref{e2.3}. 

The identification of the coefficients of
$z^kt^l$ in equations~\ref{e2.1} and~\ref{e2.3} leads to
$$
F_{k,l}={k-l\choose l-1}\,.
\label{e2.4}
$$
Using the addition/induction relation for the binomial
coefficients~[13, p 174], 
$$
{k-l+1\choose l-1}={k-l\choose l-1}+{k-l\choose l-2}
\label{e2.5}
$$
ones obtains the recursion relation:
$$
F_{k+1,l}=F_{k,l}+F_{k-1,l-1}\,.
\label{e2.6}
$$

The total number of animals with area $k$ is given by
$$
F_k=\sum_l{k-l\choose l-1}
\label{e2.7}
$$
i.e.  by the $k$th Fibonacci number (see~[13, p 303]) which
satisfies the recursion relation $F_{k+1}=F_k+F_{k-1}$ with the initial
values $F_0=0$ and $F_1=1$, according to~\ref{e2.6} and~\ref{e2.7}.

This relation with the Fibonacci numbers\footnote{$\dagger$}{A similar connection between
the number of column-convex {\it directed} animals with area $k$ and the
Fibonacci numbers with {\it odd} indices $F_{2k-1}$ has been
noticed a long time ago. See \cite{klarner65}.} can be deduced
directly by setting $t=1$ in the first line of~\ref{e2.3} which leads to 
$$
F(z,1)=F(z)=\sum_{k=1}^\infty F_k z^k={z\over 1-z-z^2}
\label{e2.8}
$$
where $F(z)$ is the generating function of the Fibonacci
numbers.

{\par\begingroup\medskip
\epsfxsize=6truecm
\topinsert
\centerline{\epsfbox{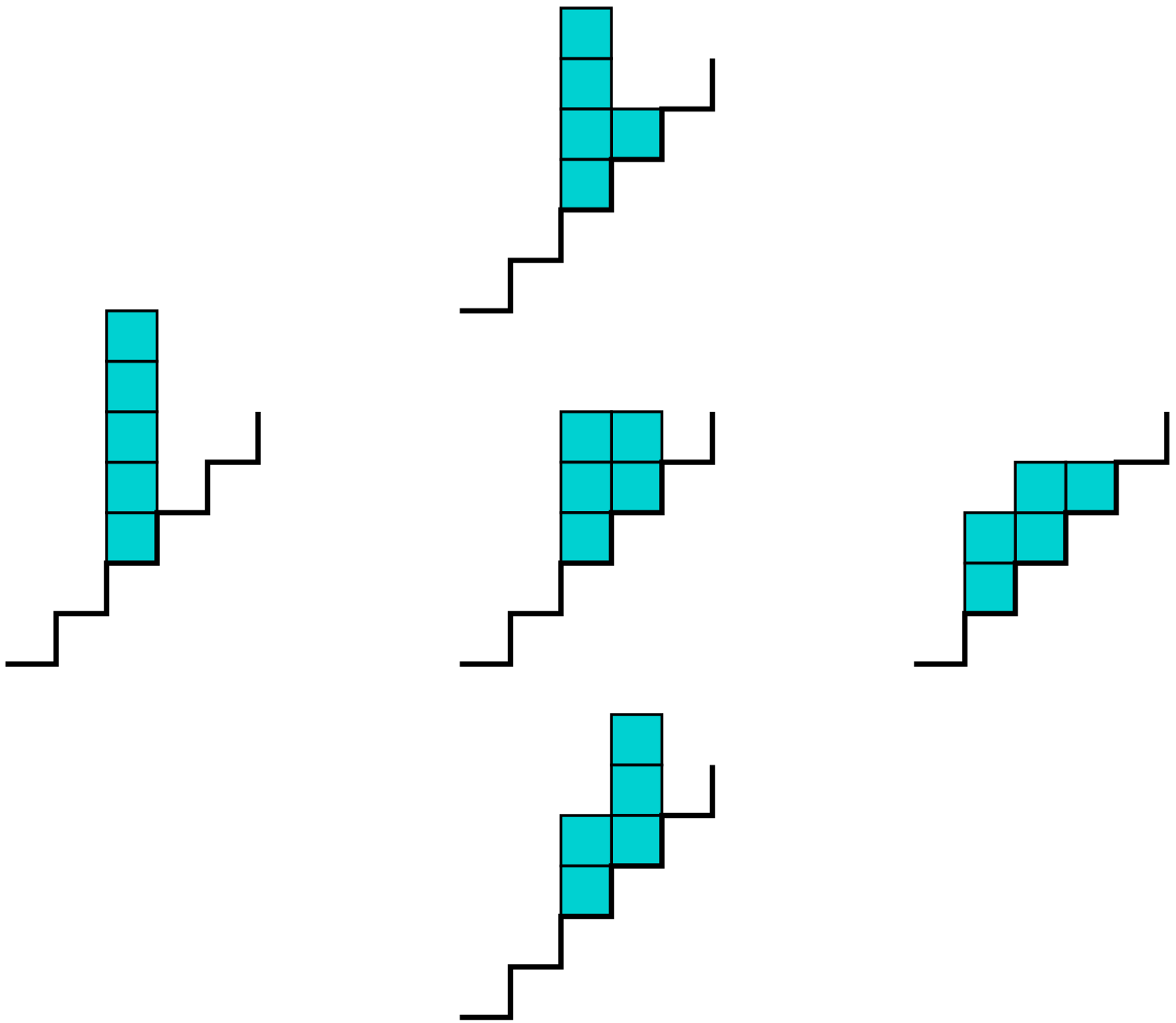}}
\vglue5mm
\figure{Different stackings of five squares on a staircase arranged by
number of occupied stairs.   
\label{fig1}} 
\endinsert
\endgroup
\par}
 
All the animals built from five squares are shown in figure~\ref{fig1}, the
different columns corresponding to the different values of $l$. 

\subsection{Mean length}
The mean length of animals with area $k$ is defined as
$$
\overline{l}(k)={\sum_{l=1}^\infty l F_{k,l}\over\sum_{l=1}^\infty
F_{k,l}}={A_k\over F_k}\qquad A_k=\sum_ll{k-l\choose l-1}\,. 
\label{e2.9}
$$
In order to calculate $A_k$, let us introduce the auxiliary generating
function 
$$
A(z)=\sum_{k=1}^\infty A_k\, z^k
=\!\left.{\partial F(z,t)\over\partial t}\right\vert_{t=1}\!\!\!\!\!
=F(z)+z F^2(z)={1-z\over z}\,F^2(z)\,\,.  
\label{e2.10}
$$
According to~[13, p 354] 
$$
F^2(z)=\sum_{k=2}^\infty F^{(2)}_k z^k\qquad 
F^{(2)}_k=\!\!\!\!\sum_{m+n=k}\!\!\!\!F_nF_m\!\!={2kF_{k+1}-(k+1)F_k\over5}
\label{e2.11}
$$
hence the coefficients of $A(z)$ are
$$
A_k\!\!=\![z^k]\,(z^{-1}\!\!-1)\,F^2(z)\!=\!F^{(2)}_{k+1}\!\!-F^{(2)}_k\!
=\!{k(2F_k\!\!-F_{k-1})+3F_k\over5}
\label{e2.12}
$$
where we used the recursion relation for the Fibonacci numbers. The
symbol $[z^k]f(z)$ means the coefficient of $z^k$ in the series
$f(z)$. The mean length follows from equations~\ref{e2.9} and~\ref{e2.12} 
and reads  
$$
\overline{l}(k)={k\over5}\left(2-{F_{k-1}\over F_k}\right)+{3\over5}\,.
\label{e2.13}
$$

The same method allows a calculation of the mean-square deviation
of the length of animals with area $k$. The generating function 
$$ 
C(z)=\sum_{k=1}^\infty C_k\, z^k\qquad 
C_k=\sum_l l(l-1)\,{k-l\choose l-1}
\label{e2.14}
$$
is given by the second derivative of $F(z,t)$ at $t=1$,
$$
C(z)=\left.{\partial^2 F(z,t)\over\partial t^2}\right\vert_{t=1}\!\!\!
=2zF^2(z)+2z^2F^3(z)\,.
\label{e2.15}
$$
The coefficients of the series $F^3(z)$
are needed to obtain $C_k$. They are calculated in the appendix with the
following result:
$$
F^3(z)=\sum_{k=3}^\infty F_k^{(3)} z^k\qquad
F_k^{(3)}={(k+1)(k+2)\over10}\, F_k-{3\over5}\, F^{(2)}_{k+1}\,.
\label{e2.16}
$$
Thus we have
$$
C_k\!=\!2F^{(2)}_{k-1}\!+\!2F^{(3)}_{k-2}
\!=\!4\!\left[{2(k-1)F_k\!-\!kF_{k-1}\over25}\right]\!+\!{k(k-1)\over5}\,F_{k-2}
\label{e2.17}
$$
and, using~\ref{e2.12},
$$
\overline{l^2}(k)=\!{C_k+A_k\over F_k}
\!=\!{1\over 25}\left[ 5k^2\left(1\!-\!{F_{k-1}\over
F_k}\right)\!+\!k\!\left(13\!-\!4\,{F_{k-1}\over F_k}\right)\!+\!7\right]\,.
\label{e2.18}
$$
The mean-square deviation $\overline{\Delta l^2}(k)
=\overline{l^2}(k)-\overline{l}^2(k)$
follows from~\ref{e2.13} and~\ref{e2.18}. 

\subsection{Mean height}
The mean height of an animal with area $k$ living on $l$ stairs
is taken as the ratio $k/l$, i.e.  it is measured from the
staircase. Thus the mean height of animals with area $k$ is given by
$$
\overline{h}(k)={\sum_{l=1}^\infty kl^{-1} F_{k,l}\over\sum_l
F_{k,l}}={B_k\over F_k}\qquad B_k=\sum_l{k\over l}\,{k-l\choose l-1}\,. 
\label{e2.19}
$$
The absorption/extraction identity can be used to write
$$
\fl
{k-l+1\choose l}={k-l+1\over l}{k-l\choose l-1}={k+1\over l}{k-l\choose
l-1}-{k-l\choose l-1}\qquad l\geq1\,,
\label{e2.20}
$$
from which we deduce:
$$
{k\over l}{k-l\choose l-1}={k\over k+1}\left[{k-l+1\choose l}+{k-l\choose
l-1}\right]\qquad l\geq1\,.
\label{e2.21}
$$
Inserting~\ref{e2.21} into the definition of $B_k$ in~\ref{e2.19}, we
obtain   
$$
\eqalign{
B_k&={k\over k+1}\left[\sum_{l\geq1}{k-l+1\choose l}+\sum_l{k-l\choose
l-1}\right]\cr
&={k\over k+1}\left[\sum_l{k-l+1\choose l}-1+\sum_l{k-l\choose
l-1}\right]\,,}
\label{e2.22}
$$
where the last equation follows from adding and substracting the term
$l=0$ in the first sum. According to the combinatorial definition of the
Fibonacci numbers in~\ref{e2.7} we have
$$
B_k={k\over k+1}\,(F_{k+2}+F_k-1)
\label{e2.23}
$$
and finally
$$
\overline{h}(k)={k\over k+1}\left({F_{k+2}\over
F_k}+1-{1\over F_k}\right)\,.  
\label{e2.24}
$$

\subsection{Asymptotic behaviour}
The generating function of the Fibonacci numbers in~\ref{e2.8} can be written
as the partial fraction expansion
$$\fl
F(z)={1\over\sqrt{5}}\left({1\over1-\phi z}-{1\over1-\hat\phi z}\right)
\qquad\phi={1+\sqrt{5}\over2}\qquad\hat\phi={1-\sqrt{5}\over2}
\label{e2.25}
$$
where $\phi\approx1.61803$ is the {\it golden ratio} and
$\hat\phi=1-\phi$. Thus the Fibonacci numbers are given by
$$
F_k={1\over\sqrt{5}}(\phi^k-\hat\phi^k)\,.
\label{e2.26}
$$

Generally the number of animals with size $k$ behaves at large size as
$Ck^{-\theta}\lambda^k$, where $C$ is a constant amplitude, $\lambda$ is the
inverse of the critical fugacity $z_c$ and $\theta$ is the exponent
governing the critical behaviour of the generating function $G(z,1)\sim
(z_c-z)^{\theta-1}$ when $z\to z_{c-}$. From~\ref{e2.7} and~\ref{e2.26} we
deduce the critical parameters: 
$$
\theta=0\qquad \lambda=z_c^{-1}=\phi\,.
\label{e2.27}
$$
According to equations~\ref{e2.13} and~\ref{e2.26}, the mean length behaves
asymptotically as
$$
\overline{l}(k)={2+\hat\phi\over5}\, k+\Or(1)
={5-\sqrt{5}\over10}\, k+\Or(1)\,.
\label{e2.28}
$$
For the mean-square length we have
$$
\overline{l^2}(k)={1+\hat\phi\over5}\, k^2+{13+4\hat\phi\over25}\,
k+\Or(1) 
\label{e2.29}
$$
whereas
$$
\overline{l}(k)^2={(2+\hat\phi)^2\over25}\,k^2+{6(2+\hat\phi)\over25}\,k+\Or(1)\,.
\label{e2.30}
$$
It is easy to verify that the leading $\Or(k^2)$ contributions
in~\ref{e2.29} and~\ref{e2.30} are the same. Thus the mean-square deviation
is $\Or(k)$ and given by:
$$
\overline{\Delta l^2}(k)={k\over5\sqrt{5}}+\Or(1)\,.
\label{e2.31}
$$

The behaviour of the mean height follows from equations~\ref{e2.24}
and~\ref{e2.26}. It tends to a constant value:  
$$
\overline{h}(k)=1+\phi^2+\Or(k^{-1})={5+\sqrt{5}\over2}+\Or(k^{-1})\,.
\label{e2.32}
$$   

\section{Double staircase}
Next we consider column-convex animals on a double staircase as shown in
figure~\ref{fig2}. The connectivity rules for neighbouring columns are the
same as above for a single staircase and we count the different stackings
with at least one square on the lowest central stair. 

\subsection{Number of animals}
The generating function for the number $G_{k,l}$ of animals with area $k$
living on $l$ successive stairs,
$$
G(z,t)=\sum_{k,l=1}^\infty G_{k,l} z^kt^l
\label{e3.1}
$$
can be expressed using the generating function for animals in a single
staircase in equation~\ref{e2.3} as 
$$
G(z,t)=zt+t\,{z^2\over1-z}\,[1+F(z,t)]^2\,.
\label{e3.2}
$$
The first term corresponds to a single square on the central stair.
When there are two or more squares on the central stair (first factor in
the second term) the animal may either stop or continue to grow on one or two
staircases (second factor in the second term). Using~\ref{e2.2}, the
generating function may be rewritten as
$$
\fl
G(z,t)=zt+[1+F(z,t)][F(z,t)-zt]=(1-zt)\,F(z,t)+F^2(z,t)\,.
\label{e3.3}
$$
Accordingly, we have
$$
\eqalign{
G_{k,l}&=[z^kt^l]F(z,t)-[z^{k-1}t^{l-1}]F(z,t)+[z^kt^l]F^2(z,t)\cr
&={k-l\choose l-1}-{k-l\choose l-2}+\sum_{p,q}{p-q\choose q-1}
{k-p-l+q\choose l-q-1}\,.
}
\label{e3.4}
$$
Using the relation [13, p 169]
$$
\sum_p{p-q\choose q-1}{k-l-p+q\choose l-q-1}={k-l+1\choose l-1}
\label{e3.5}
$$
the double sum in~\ref{e3.4} can be reduced to
$$
\fl
\sum_{q=1}^{l-1}\sum_p{p-q\choose q-1}{k-p-l+q\choose l-q-1}
=\sum_{q=1}^{l-1}{k-l+1\choose l-1}=(l-1){k-l+1\choose l-1}
\label{e3.6}
$$
yielding
$$
\fl
G_{k,l}={k-l\choose l-1}-{k-l\choose l-2}+(l-1){k-l+1\choose l-1}
=l{k-l+1\choose l-1}-2{k-l\choose l-2}
\label{e3.7}
$$
where the last expression follows from the addition/induction
relation~\ref{e2.5}.

{\par\begingroup\medskip
\epsfxsize=10.85truecm
\topinsert
\centerline{\epsfbox{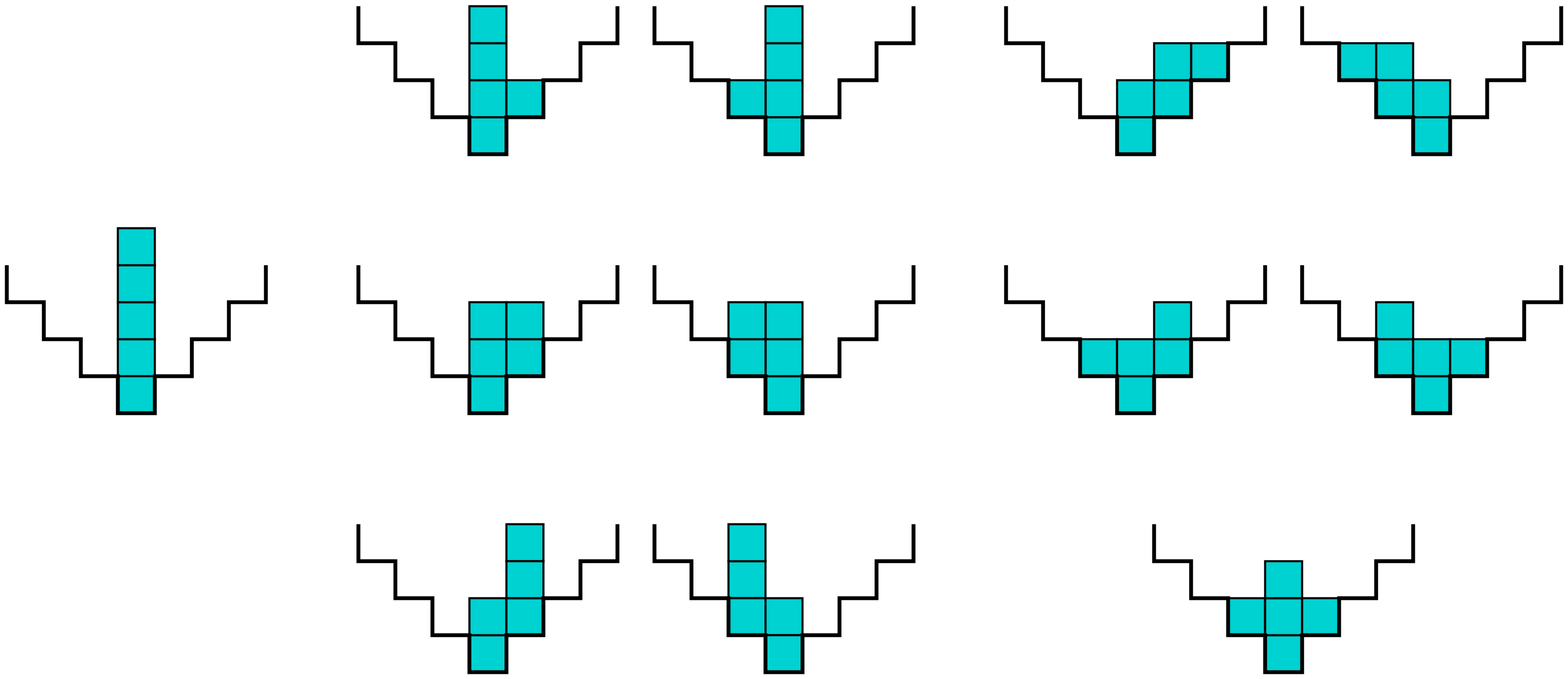}}
\vglue5mm
\figure{Different stackings of five squares on a double staircase arranged
by number of occupied stairs.   
\label{fig2}} 
\endinsert
\endgroup
\par}

The total number of animals with area $k$ reads as
$$
G_k=\sum_lG_{k,l}=\sum_ll{k-l+1\choose l-1}-2\sum_l{k-l\choose
l-2}\,.
\label{e3.8}
$$
According to~\ref{e2.9}, the first term is equal to $A_{k+1}$, whereas the
second follows from~\ref{e2.7}. Using the expression of $A_k$
in~\ref{e2.12} and the recursion on the Fibonacci numbers, we obtain
$$
G_k=A_{k+1}-2F_{k-1}={k(F_k+2F_{k-1})+4F_k-5F_{k-1}\over5}\,.
\label{e3.9}
$$

\subsection{Mean length}
As above the mean length of animals with area $k$ is defined as
$$
\fl
\overline{l}(k)={\sum_{l=1}^\infty l G_{k,l}\over\sum_{l=1}^\infty G_{k,l}}
={{\cal A}_k\over G_k}\qquad 
{\cal A}_k=\sum_l l^2{k-l+1\choose l-1}-2\sum_l l{k-l\choose l-2}\,.
\label{e3.10}
$$
The ${\cal A}_k$s are the coefficients of the generating function
$$
\eqalign{
{\cal A}(z)&=\sum_{k=1}^\infty{\cal A}_k z^k
=\left.{\partial G(z,t)\over\partial t}\right\vert_{t=1}\cr
&=-zF(z)+(1-z)\,A(z)+2F(z)A(z)\cr 
&=-zF(z)+{(1-z)^2\over z}\,F^2(z)+2\,{1-z\over z}\,F^3(z) 
}
\label{e3.11}
$$
which follows from \ref{e3.3}, taking into account the expression of $A(z)$
given in~\ref{e2.10}.

Thus, according to~\ref{e3.11}, \ref{e2.11} and~\ref{e2.16}, we have
$$
\eqalign{
{\cal A}_k&=[z^k]{\cal
A}(z)=-F_{k-1}\!\!+F^{(2)}_{k-1}\!\!-2F^{(2)}_k\!+F^{(2)}_{k+1}\!\!
+2\left(F^{(3)}_{k+1}\!\!-F^{(3)}_k\right)\cr
&={5k^2F_{k-1}+k(19F_k-7F_{k-1})+6F_k-25F_{k-1}\over25}\,. }
\label{e3.12}
$$
The mean length follows from equations~\ref{e3.9}, \ref{e3.10}
and~\ref{e3.12}
$$
\overline{l}(k)={1\over5}{5k^2F_{k-1}+k(19F_k-7F_{k-1})+6F_k-25F_{k-1}
\over k(F_k+2F_{k-1})+4F_k-5F_{k-1}}\,.
\label{e3.13}
$$
The mean-square deviation can be determined, proceeding as above for the
single staircase. It involves a lengthy calculation of the generating
function $F^4(z)$, which can be obtained in the same way as
$F^3(z)$ in the appendix. The asymptotic behaviour is the same as for the
single staircase as shown in the discussion.

\subsection{Mean height}
The mean height, measured from the staircase, takes the form
$$
\fl
\overline{h}(k)={\sum_{k=1}^\infty kl^{-1} G_{k,l}\over
\sum_{k=1}^\infty G_{k,l}}={{\cal B}_k\over G_k}\qquad
{\cal B}_k=k\sum_l{k-l+1\choose l-1}-2\sum_l{k\over l}{k-l\choose l-2}\,.
\label{e3.14}
$$
According to~\ref{e2.7}, the first sum is equal to $F_{k+1}$.
Using~\ref{e2.5} and~\ref{e2.19}, the second sum can be rewritten as
$$
\fl
\sum_l{k\over l}{k-l\choose l-2}=\sum_l{k\over l}{k-l+1\choose l-1}
-\sum_l{k\over l}{k-l\choose l-1}={k\over k+1}\,B_{k+1}-B_k\,.
\label{e3.15}
$$
Collecting these results and taking~\ref{e2.23} into account, we
obtain
$$
\eqalign{
{\cal B}_k&=kF_{k+1}+{2k\over k+1}\,(F_{k+1}+2F_k-1)
-{2k\over k+2}\,(3F_{k+1}+F_k-1)\cr
&={k\over(k+1)(k+2)}\,[k(k-1)F_{k+1}+2(k+3)F_k-2]\,.
}
\label{e3.16}
$$
Finally, the mean height follows from~\ref{e3.9} and~\ref{e3.16} and reads
$$
\overline{h}(k)={5k\over(k+1)(k+2)}\,{k^2(F_k+F_{k-1})
+k(F_k-F_{k-1})+6F_{k-2}\over k(F_k+2F_{k-1})+4F_k-5F_{k-1}}\,.
\label{e3.17}
$$

\subsection{Asymptotic behaviour}
According to~\ref{e3.9} and~\ref{e2.26}, the asymptotic
number of animals with area $k$ on a double staircase is such that
$$
\theta=-1\qquad \lambda=\phi\,.
\label{e3.18}
$$
From~\ref{e3.13} we deduce that
$$
\overline{l}(k)={k\over2+\phi}+\Or(1)={5-\sqrt{5}\over10}\,k+\Or(1)
\label{e3.19}
$$
whereas~\ref{e3.17} leads to
$$
\overline{h}(k)=5\,{1+\phi\over2+\phi}+\Or(k^{-1})
={5+\sqrt{5}\over2}+\Or(k^{-1})\,.
\label{e3.20}
$$

\section{Discussion}
The behaviour of the number of animals $F_{k,l}$ as a function of $l$ for
$k\gg1$ can be obtained by expandind $\ln F_{k,l}$ to second order near its
maximum using the Stirling approximation. This leads to
a Gaussian distribution:  
$$
F_{k,l}\simeq{5^{1/4}\phi^k\over\sqrt{2\pi k}}\,
\exp\left[{(l-\overline{l}(k))^2\over 2\overline{\Delta l^2}(k)}\right]\,.
\label{e4.1}
$$
Here $\overline{l}(k)$ and $\overline{\Delta l^2(k)}$ are the leading
contributions to the mean length and the mean-square 
deviation as given in~\ref{e2.28} and~\ref{e2.31}, respectively. The
prefactor follows from the normalization to $F_k\simeq\phi^k/\sqrt{5}$. 

The same behaviour is obtained in the double staircase for which,
according to~\ref{e3.7}, $G_{k,l}\simeq lF_{k,l}
\simeq\overline{l}(k)F_{k,l}$. Only the normalization differs
since, according to~\ref{e3.9}, $G_k\simeq k(F_k+2F_{k-1})/5
\simeq k\phi^k/5$. Hence we have
$$
G_{k,l}\simeq\sqrt{k\over2\pi\sqrt{5}}\,\phi^k
\exp\left[{(l-\overline{l}(k))^2\over2\overline{\Delta l^2}(k)}\right]\,.
\label{e4.2}
$$

Thus, to leading order, $\overline{l}(k)$ and $\overline{\Delta l^2}(k)$ are
the same  for both problems. Furthermore the Gaussian
distribution leads to $\overline{h}(k)=k\,\overline{1/l}(k)\simeq
k/\overline{l}(k)$ and we obtain the relation
$\overline{h}(k)\overline{l}(k)=k$, valid to leading order too. These
conclusions are in agreement with the exact results obtained in subsections
2.4 and 3.4.

The animals are strongly anisotropic. Their length, measured along the
staircase, scales as $\overline{l}(k)\sim k^{\nu_\parallel}$ and their
transverse size as $\overline{h}(k)\sim k^\nu$ with $\nu_\parallel=\nu z=1$
and $\nu=0$. Thus the anisotropy exponent $z$ is infinite. 

In this paper we have studied column-convex animals in staircases with
steps of unit height. The problem can be generalized by considering
staircases with steps of constant arbitrary height. Then Fibonacci numbers
are replaced by generalized Fibonacci numbers. 
 
\Appendix{Calculation of the coefficients of $F^3(z)$}
Starting from the partial fraction expansion for $F(z)$
in~\ref{e2.25} we hav: 
$$
\fl
\eqalign{
F^3(z)&={1\over5\sqrt{5}}\left({1\over1\!-\!\phi z}-{1\over1\!-\!\hat\phi
z}\right)^3\cr 
&={1\over5\sqrt{5}}\left[{1\over(1\!-\!\phi
z)^3}-{1\over(1\!-\!\hat\phi z)^3}-{3\over(1\!-\!\phi z)(1\!-\!\hat\phi z)}
\left({1\over1\!-\!\phi z}-{1\over1\!-\!\hat\phi z}\right)\right]\,. 
}   
\label{ea.1}
$$
Making use of the series expansion
$$
{1\over(1-x)^3}=\sum_k{k+2\choose k}\,x^k
=\frac{1}{2}\sum_{k=0}^\infty (k+1)(k+2)x^k
\label{ea.2}
$$
with $x=\phi z$ and $x=\hat\phi z$ in the two first terms and,
in the third, taking into account the relation 
$$
(1-\phi z)(1-\hat\phi z)=1-z-z^2
\label{ea.3}
$$
which follows from the expressions of $\phi$ and $\hat\phi$ in~\ref{e2.25},
we have
$$
F^3(z)={1\over5\sqrt{5}}\left[{1\over2}
\sum_{k=0}^\infty (k+1)(k+2)(\phi^k-\hat\phi^k)z^k
-3\sqrt{5}{F^2(z)\over z}\right]\,.
\label{ea.4}
$$
Furthermore, since
$$
\phi^k-\hat\phi^k
=[z^k]\left({1\over1-\phi z}-{1\over1-\hat\phi z}\right)=\sqrt{5}\,F_k
\label{ea.5}
$$
we can rewrite~\ref{ea.4} as
$$
F^3(z)=\sum_{k=1}^\infty {(k+1)(k+2)\over10}\, F_k\, z^k
-{3\over5}\,{F^2(z)\over z}\,.
\label{ea.6}
$$
Thus, the coefficients of the series are given by
$$
\eqalign{
F^{(3)}_k&={(k+1)(k+2)\over10}\, F_k-{3\over5}[z^k]\,{F^2(z)\over z}\cr
&={(k+1)(k+2)\over10}\, F_k-{3\over5}F^{(2)}_{k+1}\,.
}
\label{ea.7}
$$
Using~\ref{e2.11}, it is easy to verify that $F^{(3)}_1=F^{(3)}_2=0$ and 
$F^{(3)}_3=1$ as expected since the series expansion of $F(z)$ starts with
$z$.

\ack  The Laboratoire de Physique des Mat\'eriaux is Unit\'e Mixte de
Recherche CNRS No 7556.
  
\numreferences
\bibitem{bousquet96a} {Bousquet-M\'elou M 1996} {Combinatoire
\'Enum\'erative} {\it Universit\'e de Bordeaux~I Habilitation Report} p
37   
\bibitem{stauffer92} {Stauffer D and Aharony A 1992} {\it
Introduction to Percolation Theory} 2nd edition (London: Taylor and Francis)
\bibitem{bousquet96b} {Bousquet-M\'elou M 1996} {\it Eur. J. Comb.}
{\bf 17} 343  
\bibitem{klarner73} {Klarner D A and Rivest R L 1973} {\it Can. J.
Math.} {\bf 13} 585
\bibitem{nadal82} {Nadal J P, Derrida B and Vannimenus J 1982} {\JP} {\bf
43} 1561 
\bibitem{dhar82a} {Dhar D 1982} {\PRL} {\bf 49} 959
\bibitem{dhar82b} {Dhar D, Phani M K and Barma M 1982} {\JPA} {\bf 15} L279
\bibitem{hakim83} {Hakim V and Nadal J P 1983} {\JPA} {\bf 16} L213
\bibitem{brak90b} {Brak R and Guttmann A J 1990} {\JPA} {\bf 23} 4581 
\bibitem{lin91} {Lin K Y 1991} {\JPA} {\bf 24} 2411 
\bibitem{joyce94}{Joyce G S and Guttmann A J 1994} {\JPA} {\bf 27} 4359   
\bibitem{bousquet96c} {Bousquet-M\'elou M 1996} {\it Discrete Math.}
{\bf 154} 1 
\bibitem{graham94} {Graham  R L, Knuth D E and Patashnik O 1994} {\it
Concrete Mathematics: a Foundation for Computer Science} 2nd edition
(Readings, MA: Addison-Wesley)
\bibitem{klarner65} {Klarner D A 1965} {\it Fibonacci Quart.} {\bf
3} 9

\vfill\eject\bye